\documentclass[12pt]{article}
\usepackage{ifpdf}
\usepackage[english]{babel}
\usepackage{rotating}
\usepackage{caption}
\usepackage{graphicx}
\usepackage{color}
\usepackage{ulem}

\makeatletter
\renewcommand*{\@biblabel}[1]{\hfill#1.}
\makeatother

\usepackage[superscript]{cite}

\expandafter\let\csname equation*\endcsname\relax
\expandafter\let\csname endequation*\endcsname\relax
\usepackage{amsmath}
\usepackage{amssymb}
\usepackage{empheq}
\usepackage{authblk}

\definecolor{tcol}{rgb}{0.2, 0.4, 0.85}

\newcommand{\Red}{\color{black}}

\title{\textsf{\textbf{Observation of  plaquette fluctuations in the spin-1/2 honeycomb lattice}}}

\author
{
C. Wessler$^{1\ast}$, 
B. Roessli$^{1}$,
K.W. Kr\"amer$^{2}$, 
B. Delley$^{1}$,
O. Waldmann$^{3}$, 
L. Keller$^{1}$,  
D. Cheptiakov$^{1}$,
H.B. Braun$^{2,4,5}$,
and M. Kenzelmann$^{1}$,
\\
\normalsize{$^{1}$Laboratory for Neutron Scattering and Imaging, Paul Scherrer Institut, Villigen, Switzerland}\\
\normalsize{$^{2}$Department of Chemistry and Biochemistry, University of Bern, Bern, Switzerland}\\
\normalsize{$^{3}$Physikalisches Institut, Universität Freiburg, Freiburg, Germany}\\
\normalsize{$^{4}$Theoretical Physics,  ETH Z\"urich, Switzerland}\\
\normalsize{$^{5}$Dublin Institute for Advanced Studies, Dublin 4, Ireland}\\
\normalsize{$^\ast$To whom correspondence should be addressed; E-mail:  christian.wessler@psi.ch}
}

\date{}

\begin{document}

\baselineskip24pt

\maketitle 
\vspace{-1.5cm}
\begin{abstract}
\textbf{\textsf{
Quantum spin liquids are materials that feature quantum entangled spin correlations and avoid magnetic long-range order at T = 0 K.  
Particularly interesting are two-dimensional honeycomb spin lattices where a plethora of exotic quantum spin liquids have been predicted. 
Here, we experimentally study an effective S = 1/2 Heisenberg honeycomb lattice with competing nearest and next-nearest neighbor interactions. 
We demonstrate that YbBr$_{3}$ avoids order down to at least T = 100 mK and features a dynamic spin-spin correlation function with broad continuum scattering typical of quantum spin liquids near a quantum critical point.
The continuum in the spin spectrum is consistent with  plaquette type fluctuations predicted by theory. 
Our study is the experimental demonstration that strong quantum fluctuations can exist on the honeycomb lattice even in the absence of Kitaev-type interactions, and opens a new perspective on quantum spin liquids.
}}
\end{abstract}

\section*{\textsf{Introduction}}

Magnetism arises because of the quantum mechanical nature of the electron spin, yet 
for the understanding of many materials,  particularly those used in today's applications, 
a classical approach is sufficient.
Materials with strong quantum fluctuations 
are rare, but attract significant research attention since they hold enormous potential for future
 technologies \cite{Dowling2003} that make use of the long-range entanglement for quantum communication~\cite{Kitaev2003,Nayak2008}. 
Fault-tolerant quantum computers are proposed to operate with anyon quasi-particles~\cite{Kitaev2003} which exist in a class of quantum spin liquids~\cite{Kitaev2005,balents2016}.

Quantum spin liquids (QSL) {\Red are caused by quantum fluctuations which}
 reduce  the size of the ordered magnetic moment
of static magnetic structures and can {\Red affect the dynamics of the spin excitations.}
This happens in the $S=1/2$  frustrated antiferromagnetic square lattice, 
with competing nearest and next-nearest neighbour interactions, $J_1, J_2$, 
where the zone boundary spin-waves 
develop a dispersion due to the presence of quantum dimer-type fluctuations between nearest neighbors~\cite{Tsyrulin2009}. 
These fluctuations are similar to the resonant valence bond fluctuations predicted in the  frustrated  triangular lattice~\cite{Anderson1973}, 
which are believed to be relevant for high-temperature superconductors~\cite{AndersonRiceRVB}.
Frustration can be induced  by  competing interactions and depending on their relative strength,  incommensurate magnetic phases, valence bond solids with periodic ordering of local quantum states, or QSLs  with different symmetry are theoretically predicted~\cite{Kitaev2005,Mulder2010,Fouet2001,Merino2018,Wang2010,Albuquerque2011,Ganesh20132,Ferrari2019}. 
In particular, it is expected  that  frustration enforces a quantum phase transition 
at which fractionalization of magnons into deconfined spinons occurs~\cite{Senthil2004a}.

It has been a challenge to identify and understand appropriate model systems to study QSLs.
In general, lowering the dimension will increase quantum fluctuations. 
In one-dimension QSLs have been identified in antiferromagnetic (AF)  spin chains. 
Case in point are KCuF$_{3}$~\cite{Tennant1993} and Cu(C$_{6}$D$_{5}$COO)$_{2}\cdot$ 3D$_{2}$O~\cite{Dender1997}.
In two- and three-dimensions, quantum fluctuations can be enhanced by frustration, and there are several routes to achieve this: 
The inherent geometrical frustration of  kagome~\cite{Mendels2016}, triangular~\cite{Shimizu2003}, spinel~\cite{Villain1979} and pyrochlore~\cite{Canals1998} lattices may prohibit long-range ordering at low temperatures. 
Another promising candidate is the honeycomb lattice which has received relatively little attention until Kitaev's work~\cite{Kitaev2005} when it was realized that bond-dependent anisotropic interactions can stabilize a new form of QSL whose properties are known exactly. 
Representatives materials are $\alpha$-RuCl$_{3}$~\cite{Banerjee2016}, Li$_{2}$IrO$_{3}$~\cite{Singh2012} and H$_{3}$LiIr$_{2}$O$_{6}$~\cite{Kitagawa2018} which show signatures of  spin correlations due to quantum entanglement. 

Quantum fluctuations are  enhanced  in the honeycomb lattice compared to the square lattice since the number of neighbours of each spin is lower, thus placing it closer to the quantum limit. 
When next-nearest neighbour frustrating exchange interactions are sufficiently large compared to 
the nearest neighbour exchange, theories predict a quantum phase transition from a N\'eel ground state into a quantum entangled state. 
However, there is no consensus on  the nature of this ground state:  
Theories predict either a  QSL \cite{Merino2018, Wang2010} or a plaquette  valence bond crystal (pVBC)~\cite{Albuquerque2011,Ganesh20132,Ferrari2019} with different magnetic excitations which include spinons \cite{Merino2018}, 
rotons~\cite{Ferrari2019} or plaquette fluctuations\cite{GaneshPRB}.

Here, we study the static and dynamic properties of the trihalide two-dimensional compound YbBr$_{3}$ that forms a realization of the undistorted S=1/2 honeycomb lattice with frustrated interactions. 
Short-range magnetic correlations between the Yb moments develop below  $T\approx$ 3 K, but 
the correlation length is only of the order of  the size of an elementary honeycomb  plaquette 
at  $T~=$~100~mK, consistent with a  QSL ground state. 
Despite this short correlation length, 
inelastic neutron measurements reveal well defined dispersive low energy magnetic excitations close to the Brillouin zone center. At high energies and at the zone boundary, we observe  a continuum of excitations
 that we interpret as quantum fluctuations  on an elementary hexagonal plaquette.

\section*{\textsf{Results}}

\subsection*{\textsf{Crystal structure and susceptibility}}

YbBr$_{3}$ crystallizes with the BiI$_{3}$ layer structure in the rhombohedral space group $R\bar 3$ (148), where the Yb ions form perfect two-dimensional (2D) honeycomb lattices perpendicular to the 
$c$-axis, as shown in Fig. 1.
The temperature dependence of the magnetic susceptibility has a broad maximum 
around $T=3\;\mathrm{K}$,  but as shown below, there is no evidence for long-range magnetic order  
 down to at least   $T=100\;\mathrm{mK}$.
In the low temperature regime below 10 K
we observe  $\rm~\chi^{a} \approx 1.3 \chi^{c}$  which reflects a small easy-plane anisotropy. 

 The rare earth ion Yb$^{3+}$ features a $J = 7/2$ ground-state multiplet that is split by the crystal-electric field (CEF),  
 giving rise to a total of four Kramers doublets with the 
three excited CEF levels  being observable via neutron scattering. 
The first excited level is observed at $\sim15$ meV  and the ground-state doublet  is an effective $S=1/2$ state. 
From an analysis  of the measured susceptibility and the inelastic neutron data we obtain the 
CEF parameters (cf. Suppl. Info).   They  result in  ground state expectation values of 
 $ \langle J_\perp  \rangle =1.2$  and 
 $ \langle J_\parallel  \rangle =0.8$  where the subscript indicates  spin orientations measured relative to the $c$-axis.

\subsection*{\textsf{Magnetic ground state}}

 Fig.~\ref{YbBr3_Fig_2}a shows 
 the neutron diffraction pattern of the energy integrated magnetic scattering of Yb$_3$ that was determined as the 
 difference between diffraction patterns taken at  $T=100$ mK and $T= 10$ K in order 
 to eliminate the contributions of nuclear scattering. No 
 magnetic Bragg peaks are visible in the diffraction pattern, 
demonstrating  that 
 $\rm~YbBr_{3}$ avoids magnetic order down to at least this temperature.

Diffuse magnetic scattering  is centred at $(1,0,0)$ and equivalent wave-vectors,
 which implies that the short-range correlations are described by a propagation vector \textbf{Q}$_{0}$ = (0,0,0). 
Fig.~\ref{YbBr3_Fig_2}b shows the diffuse scattering as obtained from  the 2D spin wave theory described below, 
which reproduces both 
position and intensity of the observed diffuse scattering.

Fig.~\ref{YbBr3_Fig_2}c shows a cut along the $\textbf{Q} = (q,0,0)$ direction which  reveals diffuse scattering with  Lorentzian line shape that  reflects   short-range magnetic order~\cite{Collins1989}. 
From a fit to the neutron intensity  $I  \propto \kappa^2/(q^2 + \kappa^2)$, we determine an in-plane correlation length between the $\textrm{Yb}$ moments of $\xi = 1/\kappa \approx 10~\rm \AA$ at $T=100$~mK, comparable to the fourth nearest-neighbour 
distance of $10.66 ~{\rm \AA}$ which is 
$\sim$1.25 times the diameter of an Yb$_{6}$-hexagon plaquette.

\subsection*{\textsf{Magnetic excitations}}

We measured well-defined magnetic excitations at $T = 250$~mK along three cuts in the hexagonal plane. 
Within experimental  resolution we observed a single excitation branch and
no spin gap at the zone centre.
As shown in the constant energy-scans in Fig. ~\ref{YbBr3_Fig_2}d  and in Fig.~\ref{YbBr3_Fig_3}, the magnetic excitations are sharp close to the Brillouin zone center.
One of the key results of this study  is the observation of a broadening of the spectrum when the dispersion approaches the zone boundary, as shown in Fig.~\ref{YbBr3_Fig_3}. 
In fact, the inelastic neutron spectrum close to the zone boundary exhibits a continuum which extends to over twice the energy of the
well-defined  magnetic excitation.
While  low-lying excitations  are sharp, these broad excitations are only observed at higher energies.

While  it may appear surprising that we observe well defined 
excitations  even in the presence of a correlation length of merely 
 10 ${\rm \AA}$,
this agrees with  the predictions of Schwinger-Boson~\cite{Mattsson1994} and modified spin-wave~\cite{Mosadeq2016}
 theories which show that spin-waves can propagate in low-dimensional systems with short-range N\'eel order.
The well-defined excitations in YbBr$_3$ can be described by an   effective $S=1/2$ Hamiltonian including nearest and  next-nearest neighbour Heisenberg exchange coupling, and dipolar interactions between the  CEF ground state doublets,
\begin{equation}
H = -\frac{1}{2}\sum_{i, j}\sum_{\alpha, \beta}  {\cal J}_{\alpha,\beta} (i,j) 
S^{\alpha}_{i}S^{\beta}_{j},
\end{equation}
where 
 {\cal ${\cal J}_{\alpha,\beta} (i,j) =  g^2_{\alpha}   \delta_{\alpha \beta}  J(i,j)  + g_{\alpha} g_{\beta} D_{\alpha,\beta}(i,j)$},
with  $\alpha, \beta = x, y, z$  cartesian coordinates of the hexagonal cell, and 
 $S^{\alpha}_{i}$ is the $\alpha$-component of a spin-1/2 operator at site $i$. 
Here $J(i,j)$, are the exchange coupling constants 
between distinct  sites  $i$ and $j$, while $D(i,j)$ denote  the dipolar interactions. 
For the calculation of the spin wave dispersion, we use the random phase approximation (RPA) around 
the N\'eel state with spins in the hexagonal plane and $S=1/2$. 
Our measurements allow the determination of the exchange couplings, 
 while the dipolar coupling is fixed by the magnetic moment.
As shown in Fig. 3,  we find good agreement between measured and calculated spin-wave dispersions.
The nearest- and next-nearest-neighbour exchange interactions 
$J_1$, $J_2$  are obtained from a least-square fit to the data.
We obtained $ g^{2}J_{1} = -0.69(8) \, \rm{meV} $ and $ g^{2}J_{2}  = -0.09(2) \, {\rm meV} $ 
(see Suppl. Info. for details). 
We note that our  
spin wave theory does not describe all aspects of our experimental results: 
It  predicts an optical branch for 
values of the easy-plane anisotropy
that corresponds to the measured susceptibility (cf. Fig. 1), 
while we do not find experimental evidence for such a second branch. 
Also it does not explain the existence of 
an excitation continuum as we shall discuss next.

\subsection*{\textsf{Continuum of excitations}}

As shown in  Fig.~\ref{YbBr3_Fig_3}, the magnetic excitation spectrum also features weaker broad 
scattering at  energies where the optical branch is expected.
This is particularly evident near the M-points at $(0.5,0.5,0)$ and $(0.5,-1,0)$, where the excitations extend to $0.8$ - $1\;\mathrm{meV}$ and are reminiscent of scattering observed in other low-dimensional antiferromagnets~\cite{Han2012,Mourigal2013}. 
In most materials, spin-waves are long-lived excitations that are resolution-limited as a function of energy. When the spin-waves are damped or interact  with other spin-waves
 they have a finite life-time and the line-shape of the dynamical structure factor S(\textbf{Q}, $\omega$) 
broadens. We have simulated the line-shape of S(\textbf{Q}, $\omega$)  derived from our model  and convoluted it with the resolution of the spectrometer obtained from the Takin software~\cite{Weber2016} (cf. Methods). 
While the spin-wave model adequately explains the dispersion and intensity distribution close to the Brillouin zone centers,
it does not reproduce 
 the inelastic neutron line-shape close to the maximum of the dispersion of the spin-wave branch as shown in 
Fig.~\ref{YbBr3_Fig_4}.

\section*{\textsf{Discussion}}

The spin wave dispersion in YbBr$_3$ can be well described 
by a spin-1/2 Heisenberg Hamiltonian including nearest and 
next-nearest interactions with $J_2/J_1 \approx 0.13$.
This  is close to the value $J_2/J_1 \approx 0.16$~\cite{Mulder2010}, 
where classical theories predict instability of the N\'eel state, and also close to 
 $J_2/J_1 \approx 0.1$~\cite{Fouet2001,Mosadeq2016}, where 
quantum fluctuation in linear spin wave theory destroy long-range  N\'{e}el order. 
We note that other theoretical approaches find that quantum fluctuations may stabilize the N\'{e}el phase up to somewhat higher ratios of $J_2/J_1$. 
These approaches include Schwinger boson approach~\cite{Merino2018}, variational wave functions~\cite{Ferrari2017,Ferrari2019} and exact diagonalization~\cite{Albuquerque2011} which all yield a critical ratio $J_2/J_1 \approx 0.2$.
Since we do not find any evidence for static magnetism, we thus conclude that  YbBr$_{3}$ must be in close proximity of such a quantum phase transition.

In YbBr$_3$  the Yb-ion has a large magnetic  moment  of the order of  $ 2 \,\mu_B$ and therefore the  
dipolar interactions cannot be neglected. 
At the classical level, one can show that they  favour antiferromagnetic  N\'eel order with the spins along the $c$-axis~\cite{Pich1993} 
enabled by a spin gap at the zone center of $\sim$ 200~$\mu$eV.
This spin gap  caused by the  dipolar interaction is reduced by the CEF easy-plane anisotropy which contributes to a 
destabilization of the N\'{e}el state at finite temperature (Supplement Fig.~S2). 
At $g_{\rm crit} \approx  g_{zz}/g_{xx} \equiv$ 0.985 the spin gap closes and quantum fluctuations will be enhanced.  
Below that value the spins rotate into the basal plane. Linear spin wave theory predicts that easy-plane anisotropy 
entails a lifting of the degeneracy of the two spin-wave branches at the zone centre, and 
the splitting increases with increasing anisotropy. 
A large anisotropy in YbBr$_3$ would then  become measurable 
 since the branch separation becomes large enough to be resolved.
A computation of S(\textbf{Q}, $\omega$) at  $g_{\rm crit}$ is shown in Fig.~\ref{YbBr3_Fig_3} and
describes the observed dispersion and intensities of the sharp excitations very well. 

Experimentally, we have observed neither a splitting of spin waves nor a spin gap  within the available energy resolution.
This suggests that the absence of the long-range order in YbBr$_{3}$ at T  = 100 mK 
 is caused by the competition between easy-plane anisotropy and dipolar interactions that 
 accentuates quantum fluctuations.
This places YbBr$_{3}$ close to the quantum critical point towards a QSL 
 of the spin-1/2 Heisenberg Hamiltonian on the honeycomb lattice.

Our experiment provides clear evidence for the presence of a continuum of excitations at high energies in YbBr$_{3}$.
We can exclude the possibility of the line-shape broadening being caused by two-magnon decay.
The necessary cubic anharmonicities are absent for collinear magnets such as  YbBr$_3$~\cite{Zhitomirsky2013}.
We observe that the intensity of the continuum is stronger at the M\rq{} points along (h,-1,0) and (h,h,0) directions 
whereas it is absent along (0,k,0)  and  at the $\Gamma$ and $\Gamma\rq{}$ points.

We found, as shown in Fig.~\ref{YbBr3_Fig_5}, that this modulation of the neutron intensity associated with the  continuum 
can be reproduced by a  random-phase approximation (RPA) calculation for a hexamer plaquette with the exchange parameters obtained from the spin-wave calculations  (cf. Methods). 
This picture of local excitations in YbBr$_{3}$ is supported by an analogous calculation of the magnetic susceptibility which shows a broad maximum at T $\simeq$ 4 K (see Supplement). Similar excitations associated with small spin clusters were also observed in the spinel lattice~\cite{Lee2002}.
Our neutron measurements  are also  in agreement with recent Monte-Carlo calculations of the dynamical structure 
factor for the frustrated honeycomb lattice~\cite{Ferrari2019}  that show a  deconfined two-spinon
continuum~\cite{Ferrari2018} with enhanced intensity at the zone boundary due to 
proximity of a quantum critical point.

In summary, we have shown that the magnetic ground-state of YbBr$_3$ 
exhibits only short range order
well below the maximum in the static susceptibility. 
Analysis of the dispersion of the magnetic excitations reveals competition between the nearest-neighbour and next-nearest-neighbour exchange interactions, 
but no mode softening 
We observed a continuum of excitations 
with the 
spectrum of excitations extending
 to approximately twice the energy of the position of the maximum in S(\textbf{Q}, $\omega$).
The neutron inelastic intensity  due to the continuum follows the modulation expected for the fluctuations 
of a honeycomb spin plaquette. 
Our results demonstrate that YbBr$_3$ is  a two-dimensional  $S=1/2$ system on the honeycomb lattice with spin-liquid properties without Kitaev-type interactions. 
The observation of the continuum associated with localized plaquette  excitations supports the view of a deconfined quantum critical point \cite{Senthil22004} in the frustrated honeycomb lattice, in agreement with results from coupled cluster methods, density matrix renormalization group calculations and Monte-Carlo simulations \cite{Bishop2013,Ganesh20132, Ferrari2019}.
Our measurements set a quantitative benchmark for future theoretical work.

\clearpage
\section*{\textsf{Methods}}

\subsection*{\textsf{Experimental methods}}

In the following we describe the different methods used to obtain the results presented in the main text.

\subsubsection*{\textsf{Crystal growth and sample preparation}}
An YbBr$_{3}$ single crystal of cylindrical shape (15 mm diameter, 18 mm height) was grown from the melt in a sealed silica ampoule by the Bridgman method, as previously described for ErBr$_{3}$~\cite{Kraemer1999}. 
YbBr$_{3}$ was prepared from Yb$_{2}$O$_{3}$ (6N, Metall Rare Earth Ltd.) by the NH$_{4}$Br method~\cite{Meyer1994} and sublimed for purification. 
All handling of the hygroscopic material was done under dry and O-free conditions in glove boxes or closed containers.

\subsubsection*{\textsf{Magnetic Susceptibility}}
The magnetic susceptibility was determined with a MPMS SQUID system (Quantum Design).

\subsubsection*{\textsf{Neutron scattering experiments}}
The neutron experiments were performed at the Swiss Spallation Neutron Source (SINQ) utilizing different instruments. 
On all instruments filters were used to reduce contamination of the beam by higher-order neutron wavelengths.

\noindent
\textit{Powder diffraction---}
The crystal structure of YbBr$_3$ was refined using diffraction data collected with the high-resolution powder diffractometer HRPT
at the wavelength of $\rm \lambda = 1.494~\AA$ at room temperature.
The crystal structure and lattice parameters were refined with Fullprof. 

\noindent
\textit{Diffuse scattering---}
The magnetic ground-state was investigated with the multi-counter diffractometer DMC at the wavelength $\rm \lambda = 2.4576~\AA$. 
The measured neutron intensity is proportional to the equal time spin-spin correlation function.

\noindent
\textit{Crystal-field excitations---}
The crystal-field splitting of the $\rm Yb^{3+}$ ions was determined on the thermal three-axis spectrometer EIGER operated in the constant final-energy mode with $k_f=2.662~{\rm \AA}^{-1}$ 
at $T$ = 1.5 K  and $|{\bf Q}|=1.5 \;{\rm \AA}^{-1}$.
With that configuration the energy resolution is $0.8~{\rm meV}$.

\noindent
\textit{Magnetic excitations---}
The dispersion of the magnon excitations is bound by $\rm \hbar \omega$(\textbf{q}) $\rm < 1~meV$ in YbBr$_3$ which required the use of cold neutrons that provide an improved energy resolution. 
Therefore the measurements of the spin-waves were performed with the TASP three-axis spectrometer using  $k_f=1.3~{\rm \AA}^{-1}$ which resulted in an energy-resolution of $ 80~\mu {\rm eV}$. 
To maximize the intensity, the measurements were performed without collimators in the beam and the analyzer was horizontally focusing.

\subsection*{\textsf{Theoretical methods}}

\subsubsection*{\textsf{Magnetic excitations}}

We analyzed the dispersion of the magnetic excitations with a Heisenberg Hamiltonian, 
\begin{equation}
H_{\rm h} = -\frac{1}{2}\sum_{i, j}  \sum_{\alpha} g^2_{\alpha}  \; J(i,j) \;
S^{\alpha}_iS^{\alpha}_j.
\label{Hh}
\end{equation}
$J(i,j)$ are the exchange constants between sites $i$ and $j$, to be determined experimentally, the anisotropic $g$-factors reflect the cystal-field anisotropy where $\alpha = x,y,z$ denotes Cartesian coordinates, 
 and  $S^{\alpha}_i$  denotes the $\alpha$ component of a spin-1/2 operator at site $i$.
For  Heisenberg interactions $g_{x}=g_{y}=g_{z}\equiv g$.
Because the magnetic moment of Yb$^{3+}$ is large, we also consider the dipolar interactions, 
\begin{equation}
H_{\rm dip} = -\frac{\mu_0 \mu_B^2}{8\pi}   \sum_{i,j}\sum_{\alpha,\beta}
g_{\alpha}g_{\beta}
D_{\alpha,\beta}(ij) S_{i} ^{\alpha}  S_{j} ^{\beta},
\label{Hdip}
\end{equation}
with 
\begin{equation}
D_{\alpha,\beta}(ij)=\frac{3(R_{ij})_{\alpha} (R_{ij})_{\beta} }{
R_{ij}^5}
-\frac{1}
{R^3_{ij}}  \delta_{\alpha,\beta}.
\end{equation}
where  ${\bf R}_{ij} \equiv {\bf R}_{j}  - {\bf R}_{i} $ 
is the relative position vector between the $j$'th and $i$'th  ion.

The dispersion of magnetic excitations was calculated within the random-phase approximation (RPA) 
where the spin-waves appear as poles in the dynamical tensor $\overline{\overline \chi}(\textbf{q},\omega) $,
\begin{equation}
\overline{\overline \chi}(\textbf{q},\omega) = [\overline{\overline 1} -\overline{\overline \chi}_0(\omega)\overline{\overline M}(\textbf{q})]^{-1}\overline{\overline \chi}_0(\omega)
\label{dynamsusc}
\end{equation}
 with $\overline{\overline M}(\textbf{q})$ the Fourier transform of the exchange and dipolar interactions and $\overline{\overline \chi}_0(\omega)$ the single-ion susceptibility. 
 The neutron cross-section is proportional to the imaginary part of the dynamical susceptibility~\cite{Jensen1991}, 

\begin{equation}
{d^2 \sigma \over d\Omega d E} \propto 
 \sum_{\alpha,\beta}(\delta_{\alpha,\beta}-\frac{Q_\alpha Q_\beta}{|\textbf{Q}|^2}) \; 
 S^{\alpha,\beta}(\textbf{Q}, \omega),
\label{ncs}
\end{equation}
where we defined the dynamical structure factor, 
\begin{equation}
S^{\alpha,\beta}(\textbf{Q}, \omega) = {1 \over \pi}
 \frac{1}{1-\exp(-\hbar\omega/k_BT)}  \sum_{u,v} \; \Im\chi^{\alpha,\beta}_{u,v}(\textbf{Q}, \omega).
\label{Sdef}
\end{equation}
Here  \textbf{Q} denotes the scattering vector, and $u,v$ labels the Yb-ions in the magnetic cell. 
To analyse the data, the scattering cross-section was convoluted with the resolution of the spectrometer using Popovici method implemented in Takin~\cite{Weber2016}.

\clearpage

\section*{\textsf{References}}
\begingroup
\renewcommand{\section}[2]{}

\endgroup

\clearpage
\section*{\textsf{Acknowledgments}}
\textbf{Funding:} 
The financial support by the Swiss National Science Foundation under grant no. SNF 200020$\_$172659 is gratefully acknowledged.
\textbf{Author contributions:}
D.C., L.K., M.K., B.R. and C.W. performed the neutron experiments. 
Crystal growth and characterization was done by K.W.K..
Theoretical calculations were performed by  B.D.,  B.R., C.W., O.W., and H.B.B.. 
Data analysis and discussion of the results was done by all authors. 
All authors contributed to the writing of the manuscript.
\textbf{Competing interests:}
The authors declare no competing interests.
\textbf{Data and materials availability: } 
All data needed to evaluate the conclusions in the paper are present in the paper and/or the Supplementary Materials. 
Additional data related to this paper may be requested from the authors.

\clearpage
\section*{\textsf{Figure 1}}
\begin{figure}[ht]
\begin{center}
\includegraphics*[width =\textwidth]{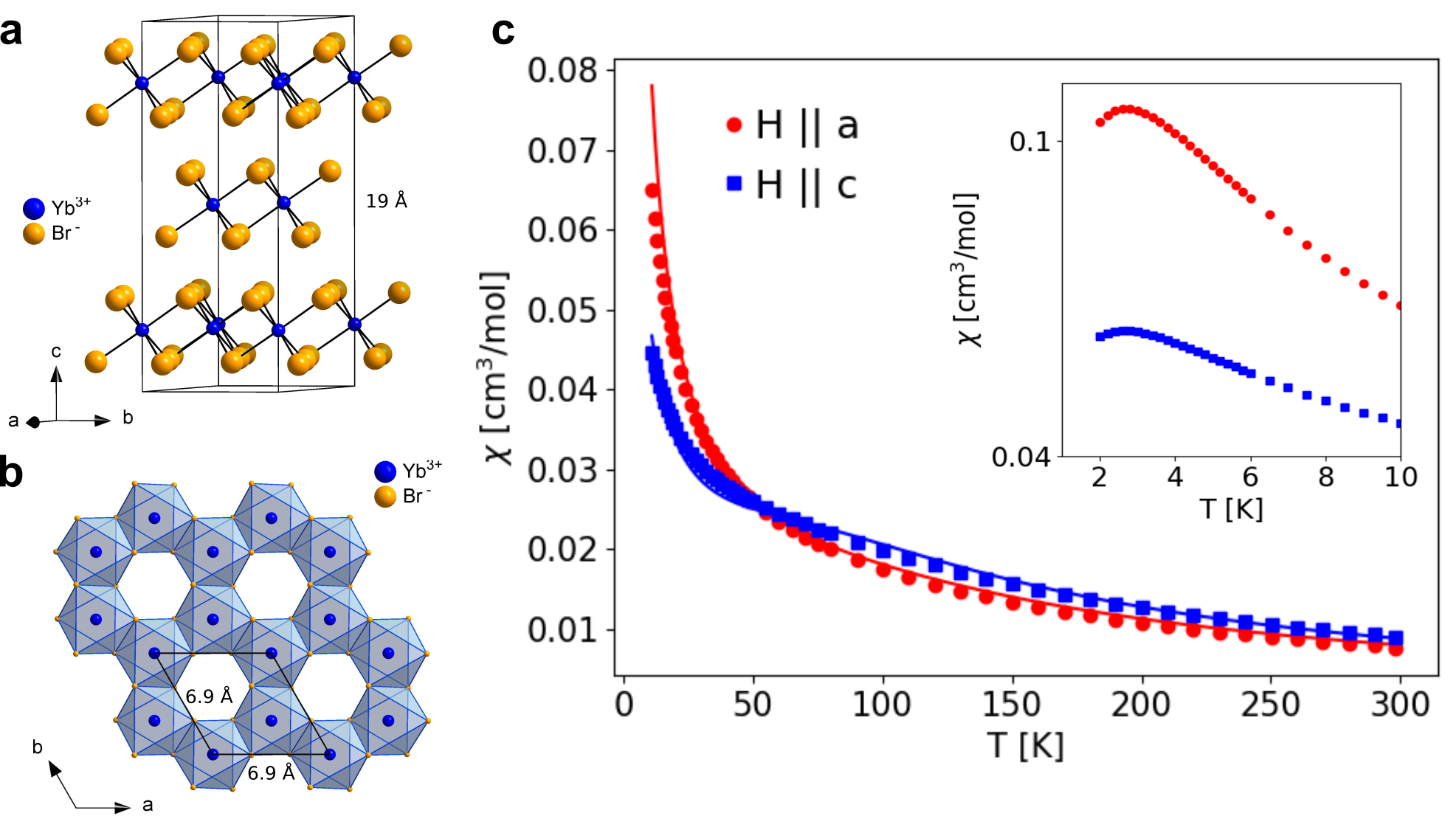}
\captionsetup{labelformat=empty}
\caption{ \textbf{Fig.1 $\vert$ Magnetic susceptibility, crystal electric field and crystal structure.}
\textbf{a}, View along [210] on the unit cell of YbBr$_3$.
\textbf{b}, Yb$^{3+}$ honeycomb layer.  
\textbf{c}, Temperature dependence of the  magnetic susceptibility $\chi$ of YbBr$_3$ for field orientations along the a- and c-axes. 
Solid lines are the calculated single-ion susceptibilities based on the crystal field (CEF)-parameters (Suppl.Info.).
Inset: Measured low-temperature susceptibility showing a rounded peak  
around  $T=2.75 \; {\rm K}$. 
 }
\label{YbBr3_Fig_1}
\end{center}
\end{figure}

\clearpage
\section*{\textsf{Figure 2}}
\begin{figure}[ht]
\begin{center}
\vspace{-0.9cm}
\includegraphics*[width =0.9\linewidth]{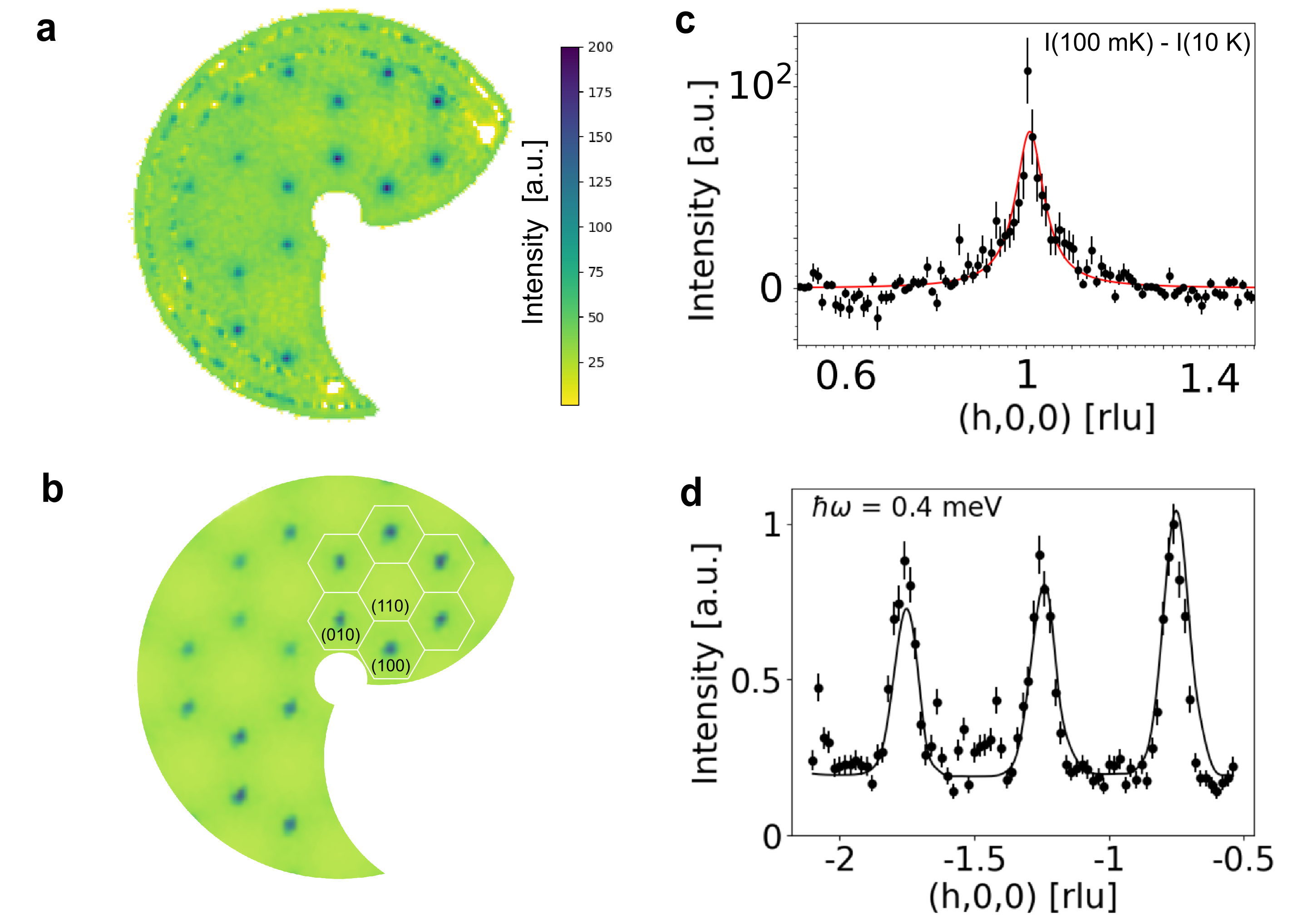}
\captionsetup{labelformat=empty}
\caption{\textbf{Fig.2 $\vert$ Magnetic diffuse scattering and correlation length.}
\textbf{a}, Magnetic diffuse scattering in YbBr$_3$  in the [h,k,0] plane at T = 100~mK, after subtraction of the nuclear Bragg contribution. 
 \textbf{b}, Calculated magnetic diffuse scattering based on the spin-wave model including exchange and dipolar interactions.
\textbf{c}, Cut through the diffuse scattering along the (h,0,0) direction. 
The line is a fit to the data with a Lorentzian function. 
[Note that the presence of paramagnetic scattering at 10 K leads to a negative background
in the 100 mK data after subtraction].
%
\textbf{d}, Constant-energy scan for $\rm \hbar \omega = 0.4~meV$ in YbBr$_3$ at T = 250~mK 
showing well-defined low energy excitations. 
The solid line represents the computed inelastic neutron scattering cross-section. 
Observed small peaks are due to
spurious scattering and are not included in the model calculation.
}
\label{YbBr3_Fig_2}
\end{center}
\end{figure}

\clearpage
\section*{\textsf{Figure 3}}
\begin{figure}[ht]
\begin{center}
\includegraphics*[width =0.9\linewidth]{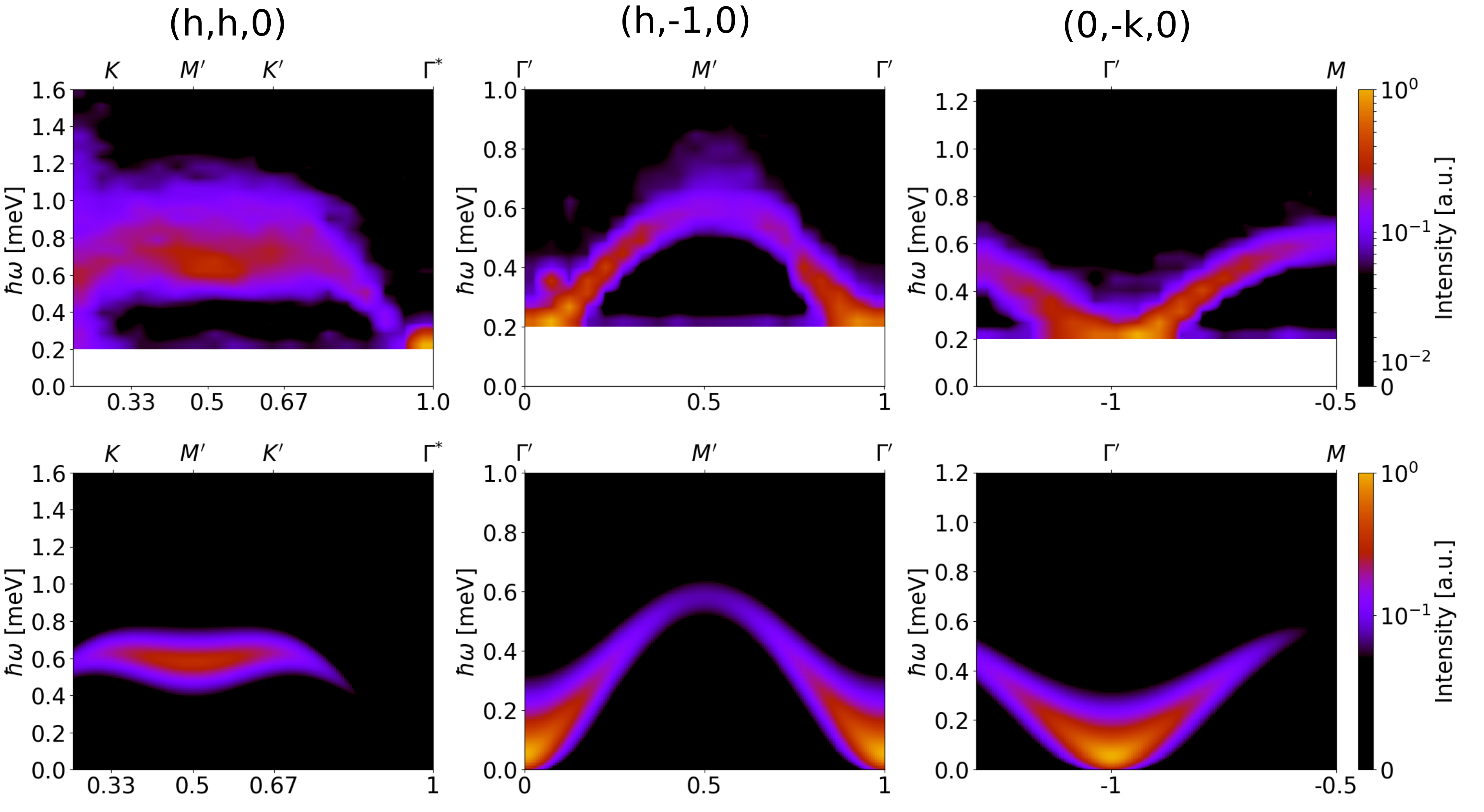}
\captionsetup{labelformat=empty}
\caption{\textbf{Fig.3 $\vert$ Magnetic excitations along high symmetry directions.}
False color plot of the observed (top) and calculated (bottom) inelastic neutron cross section 
 of the magnetic excitations in YbBr$_3$ at $T = 0.25$ K. 
The intensity is shown on a logarithmic scale. 
Note the existence of a continuum of excitations around $(1/2,1/2,0)$ and $(1/2,-1,0)$ which is not described 
by spin waves and is indicative of plaquette fluctuations (cf. Fig. 5).}
\label{YbBr3_Fig_3}
\end{center}
\end{figure}

\clearpage
\section*{\textsf{Figure 4}}
\begin{figure}[ht]
\begin{center}
\includegraphics*[width =10cm]{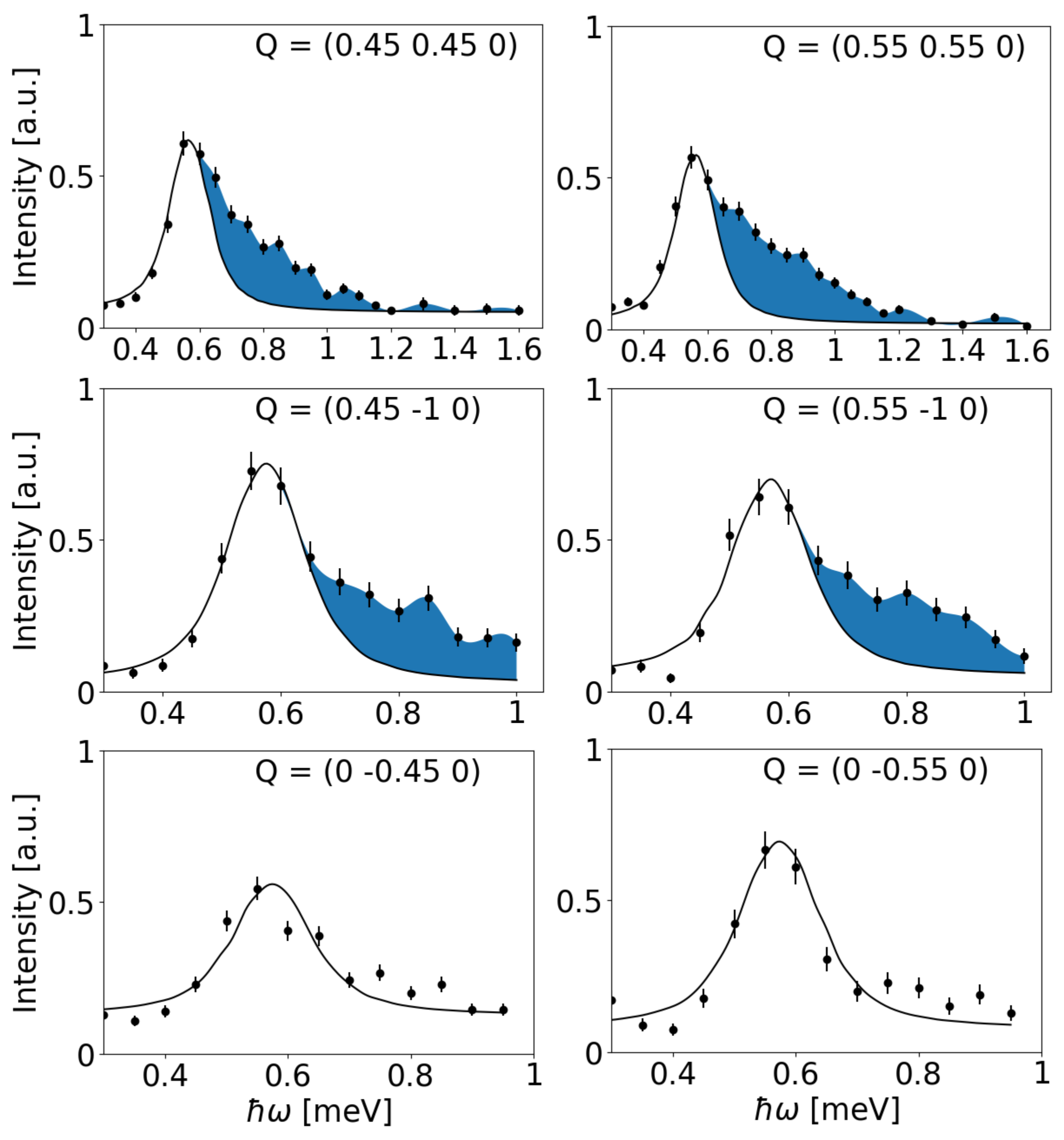}
\captionsetup{labelformat=empty}
\caption{
\textbf{Fig.4 $\vert$ Excitation continuum near the Brillouin zone boundary.} 
Observed and simulated magnon spectra based on the spin-wave model explained in the text. 
The lines are fits to the data with a Lorentzian function. An intrinsic line-width of $ 0.1~{\rm meV}$ was used for the simulation. 
The shaded area indicates the continuum of excitations.
}
\label{YbBr3_Fig_4}
\end{center}
\end{figure}

\clearpage
\section*{\textsf{Figure 5}}
\begin{figure}[ht]
\begin{center}
\includegraphics*[width =\linewidth]{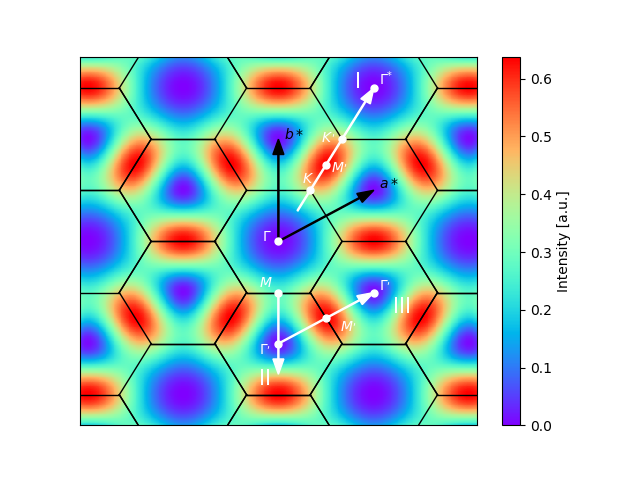}
\captionsetup{labelformat=empty}
\caption{\textbf{Fig.5 $\vert$ Calculated neutron form factor for a plaquette.} 
The neutron form  factor of an Yb$_6$ hexagon 
calculated within the random-phase approximation 
and assuming N\'eel order on the plaquette. 
The directions of the neutron measurement are indicated by white arrows:
cut I corresponds to  $(h,h,0)$, cut II is $(0,-k,0)$ and  cut III is $(h,-1,0)$.
High-symmetry points  are  labeled similar to Fig.~\ref{YbBr3_Fig_3}.
Basis vectors of the reciprocal lattice are denoted as ${\bf a^*}, {\bf b^*}.$
}
\label{YbBr3_Fig_5}
\end{center}
\end{figure}

\clearpage
\section*{\textsf{Supplementary materials}}

The supplementary materials include
\begin{itemize}
\item Section S1: Crystal structure
\item Section S2: Magnetic susceptibility
\item Section S3: Crystal electric field
\item Section S4: Magnetic ground-state
\item Section S5: Magnetic excitations
\item Table 1: Structural parameters of YbBr$_{3}$ determined on HRPT at room temperature
\item Fig. S1: Crystal electric field and magnetic susceptibility fit for YbBr$_3$
\item Fig. S2: Dependence of the energy gap as a function of easy-plane anisotropy
\item Fig. S3: Calculated susceptibility for a Yb$_{6}$ hexamer
\item References
\end{itemize}

\subsection*{\textsf{Section S1: Crystal structure}}
YbBr$_{3}$ crystallizes with the BiI$_{3}$ layer structure in the rhombohedral space group $R\bar3$ (148) with lattices parameters of $a$ = 6.97179(18)$\rm~\AA$ and $c$ = 19.1037(7)$\rm~\AA$  at room temperature.
The lattice parameters are in good agreement with powder~\cite{PDF2} and crystal~\cite{icsd} diffraction data found in literature. 
The unit cell contains six Yb$^{3+}$ ions on site (6c) at (0, 0, z), (0, 0, z) + (2/3, 1/3, 1/3) and (0, 0, z) + (1/3, 2/3, 2/3) with z = 0.1670(2). 
The Yb ions have C$_{3}$ point symmetry and form two-dimensional (2D) honeycomb lattices perpendicular to the $c$-axis, see Fig. 1.
Yb$^{3+}$ has a distorted octahedral coordination by Br$^{-}$ ions which are located on site (18f) at (x, y, z) with x=0.3331(5), y=0.3131(5), and z=0.08336(15).
Surprisingly, the distance between Yb$^{3+}$-Br$^{-}$ varies by less than  $10^{-2}\, {\rm \AA}$, however the Br$^{-}$-Yb$^{3+}$-Br$^{-}$ bond angles differ significantly between 87.3$^{\circ}$ and 91.1$^{\circ}$.  The crystallographic parameters determined on HRPT are are 
summarized in Table 1.

\subsection*{\textsf{Section S2: Magnetic Susceptibility}}
The temperature dependence of the static susceptibility $\chi$ is shown in Fig.~\ref{YbBr3_Fig_1}c for magnetic field orientations in-plane ($a$-axis) and out-of-plane ($c$-axis). 
$\chi T$ values (not shown) increase with temperature and do not saturate up to 300~K. 
The values at 300~K are 2.282 and 2.687~cm$^{3}$K/mol along the a- and c-axes, respectively. 
The average of 2.417~cm$^{3}$K/mol is slightly below the expectation value of 2.572~cm$^{3}$K/mol for the $^{2}F_{7/2}$ ground-state of Yb$^{3+}$. 
At lower temperature a maximum in the $\chi$ versus $T$ curves is observed at $T = 2.75~{\rm K}$, see the inset in Fig.~\ref{YbBr3_Fig_1}c.
We have calculated the temperature dependence of the static susceptibility for an Yb$_{6}$ honeycomb 
with the exchange parameters determined from the spin-wave analysis and easy-plane anisotropy parameters g$_a$/g$_c$ = 1.25. 
We find that the susceptibility has a broad maximum around T $\simeq$ 4 K and reproduces the experimental $\chi$ (T) 
above  5 K well, as shown in Fig. S2.

\subsection*{\textsf{Section S3: Crystal electric field (CEF)}}
The electrostatic  potential originating from the ions surrounding the Yb$^{3+}$ ion
can be modelled with Stevens operators
\begin{align*}
\begin{aligned}
H_{CEF} = \sum_{l,m}B_l^mO_l^m
\end{aligned}
\end{align*}
with $B_l^m= \gamma_l^m \theta_l$ and $\theta_l$ the Stevens coefficients.
For the $C_3$ point group symmetry of the Yb site only the parameters \cite{Bauer09}
$B_2^0,B_4^0,B_4^{\pm 3},B_6^0,B_6^{\pm 3},B_6^{\pm 6}$ are non-zero.
From the inelastic neutron scattering measurement we determined 3 CEF excitations at
$E_1=14.5$~meV, $E_2=25$~meV and $E_3=29$~meV. 
We first used the susceptibility $\chi(T)$ for the
determination of the CEF Hamiltonian. From a least-square fit to $\chi_a$ and
$\chi_c$ where $a$ and $c$ denote the crystallographic axis we obtain
$\gamma_2^0$ = -5.14~meV,
$\gamma_4^0$ = -0.59~meV,
$\gamma_4^{+3}$ = 57.43~meV,
$\gamma_4^{-3}$ = 51.31~meV,
$\gamma_6^0$ = 6.09~meV,
$\gamma_6^{+3}$ = 50.21~meV,
$\gamma_6^{-3}$ = 55.56~meV,
$\gamma_6^{+6}$ = 33.9~meV,
$\gamma_6^{-6}$ = 42.4~meV.
In agreement with the Kramers theorem the CEF splits the $J=7/2$ multiplet  of 
the Yb$^{3+}$ ion into 4 doublets. The calculated CEF-levels are at
$15.16$~meV, $24.75$~meV, and $28.88$~meV, respectively.
From a subsequent  fit of the inelastic neutron data we obtain very similar values, 
$\gamma_2^0$ =-6.49~meV,
$\gamma_4^0$ = -0.51~meV,
$\gamma_4^{+3}$ =58.53~meV,
$\gamma_4^{-3}$ =52.12~meV,
$\gamma_6^0$ = 6.01~meV,
$\gamma_6^{+3}$ = 48.11~meV,
$\gamma_6^{-3}$ = 56.30~meV,
$\gamma_6^{+6}$ = 33.21~meV,
$\gamma_6^{-6}$ = 41.12~meV. 
We show in  Fig.~S1 a comparison between calculated and observed 
neutron scattering intensities. 
We point out that first excitated CEF-level
has a double-peak structure in YbBr$_3$ that is not explained by our
model. 
It is conceivable that this peak structure is caused by a phonon, and as it 
also resembles the CEF levels\cite{Sala19} of YbCl$_3$ this issue requires further
investigation.
Nevertheless the CEF-model presented here provides an
adequate description of the temperature dependence of the static
susceptibility.
In addition we performed a point charge calculation 
based on 
the program multiX~\cite{Uldry2012}. 
In agreement with the susceptibility measurements, calculations show that at high temperatures anisotropy is small in YbBr$_{3}$ with easy-plane anisotropy developing below T = 50~K. 
At T = 4~K, we obtain   $\chi_a \approx 1.3\chi_c$.

\subsection*{\textsf{Section S4: Magnetic ground-state}}
In mean field theory, the classical ground-state is given by the eigenvectors of the largest eigenvalue $\lambda$ (\textbf{q}) of the Fourier transform of the interaction matrix $\overline{\overline M}$(\textbf{q})~\cite{Reimers1992,Kadowaki2002,Enjalran2004}. 
Based on the Hamiltonian $H_{\rm h}+H_{\rm dip}$, 
 $\lambda$ (\textbf{q}) has a maximum at \textbf{Q}$_0$ = (0,0,0) which agrees with the diffuse scattering observed in 
 YbBr$_3$  (cf. Fig. 2a).  We 
find that the dipolar energy becomes independent of the distance between the Yb-planes for a lattice parameter $c > 20~{\rm \AA}$, which shows that the 2D limit is reached in YbBr$_3$ and inter-layer interactions can be neglected.

\subsection*{\textsf{Section S5: Magnetic excitations}}
  
Because of the large separation between the ground-state and the first CEF doublet, the magnetic properties of  YbBr$_3$ can be approximated by a spin $S=1/2$. Choosing a local coordinate frame with the $\zeta$-axis oriented along a given spin direction, 
the non-zero elements of the single-ion susceptibility matrix are $\chi_{0}^{\xi\xi}(\omega)=\chi_{0}^{\eta\eta}(\omega)$ and $\chi_{0}^{\xi\eta}(\omega)=-\chi_{0}^{\eta\xi}(\omega)$ which correspond to excitations transverse to the (local) spin direction. 
Within mean-field approximation, 
\begin{eqnarray}
\chi^{\xi\xi}_0(\omega)  &=& {1 \over 2} \frac{\Delta}{\Delta^2 -(\omega +i\epsilon)^2} \\
\chi_0^{\xi\eta}(\omega)  &=& { i \over 2}\frac{\omega+i\epsilon}{\Delta^2 -(\omega +i\epsilon)^2},
\label{chi0eq}
\end{eqnarray}
with $\Delta \equiv  \Delta_i = -\langle S_\zeta \rangle \sum_{j} {\cal J}_{\zeta \zeta} (i,j)$ 
the local field acting on a given  Yb moment with ${\cal J}$ defined in 
Eq. (1) and $\epsilon$ the finite line width of the excitations.

Within linear spin-wave theory, the dipole-dipole interactions induce a gap in the spin-wave dispersion~\cite{Pich1993,Pich1995}. 
With  a Yb magnetic moment of $\rm 2~\mu_{ B}$, the dipolar interactions produce a spin gap at the zone center $\rm \sim 200 ~ \mu eV$. 
The easy-plane anisotropy favors alignment of the spins in the hexagonal plane. 
The spin gap opened by $H_{\rm dip}$ is reduced by the easy-plane anisotropy. 
At $g_{\rm crit}\sim 0.985$ the spin gap is minimal and below that value the spins rotate into the basal plane, see Fig.~S2. 
{\Red The  easy-plane anisotropy lifts the degeneracy of the spin wave branches at the 
zone center and the splitting increases with increasing anisotropy. }

\clearpage
\section*{\textsf{Supp. Tables}}

\subsection*{\textsf{Table 1: Structural parameters of YbBr$_{3}$ determined on HRPT at room temperature}}
\begin{table}[h]
\begin{tabular}{| c || c | c | c  | c | c  | c  |}
	\hline
	Name & x & y & z   &  occ. \\ \hline\hline
	Yb1 & 0 & 0 & 0.33289(21)  & 0.317(3)  \\ \hline
	Yb2 & 0 & 0& 0  & 0.009(0) \\ \hline
	Br & 0.35362(57) & 0.00022(60) & 0.08325(15)  & 1 \\ \hline
\end{tabular}
\label{SuppTable}
\end{table}

\clearpage
\section*{\textsf{Supp. Figures}}

\subsection*{\textsf{S1:  Crystal electric field.}}
	
\begin{figure}[hb]
\begin{center}
\includegraphics*[width =12cm]{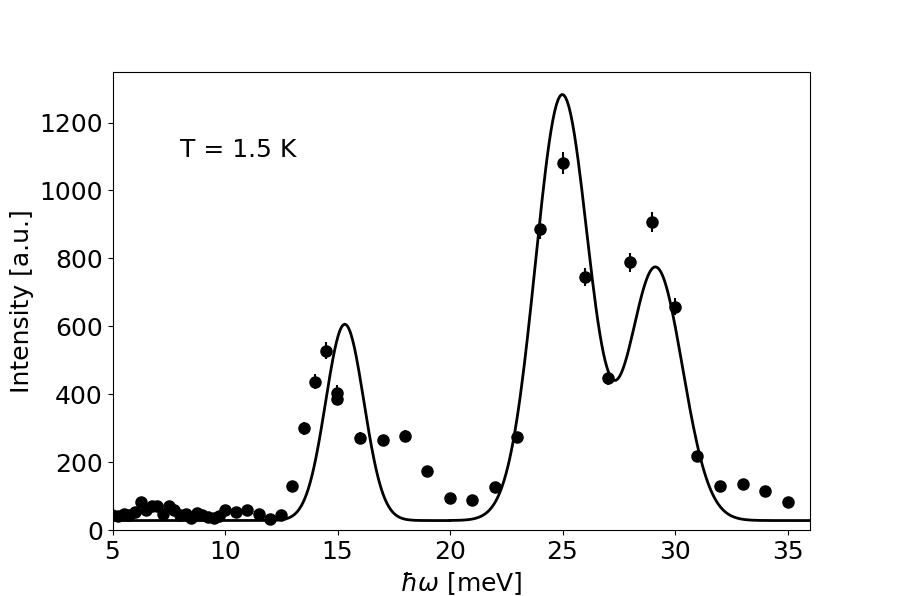}
\captionsetup{labelformat=empty}
\caption{
	\textbf{Fig. S1 $\vert$ Yb$^{3+}$ crystal electric field in YbBr$_3$. } 
Crystal electric field (CEF) excitations measured by inelastic neutron scattering. 
The solid line is the calculated intensity for the CEF-parameters given in Suppl. Info. S3.
 }
\label{YbBr3_CEF}
\end{center}
\end{figure}

\clearpage
\subsection*{\textsf{S2:  Dependence of the energy gap as a function of easy-plane anisotropy.}}
	
\begin{figure}[hb]
\begin{center}
\includegraphics*[width =12cm]{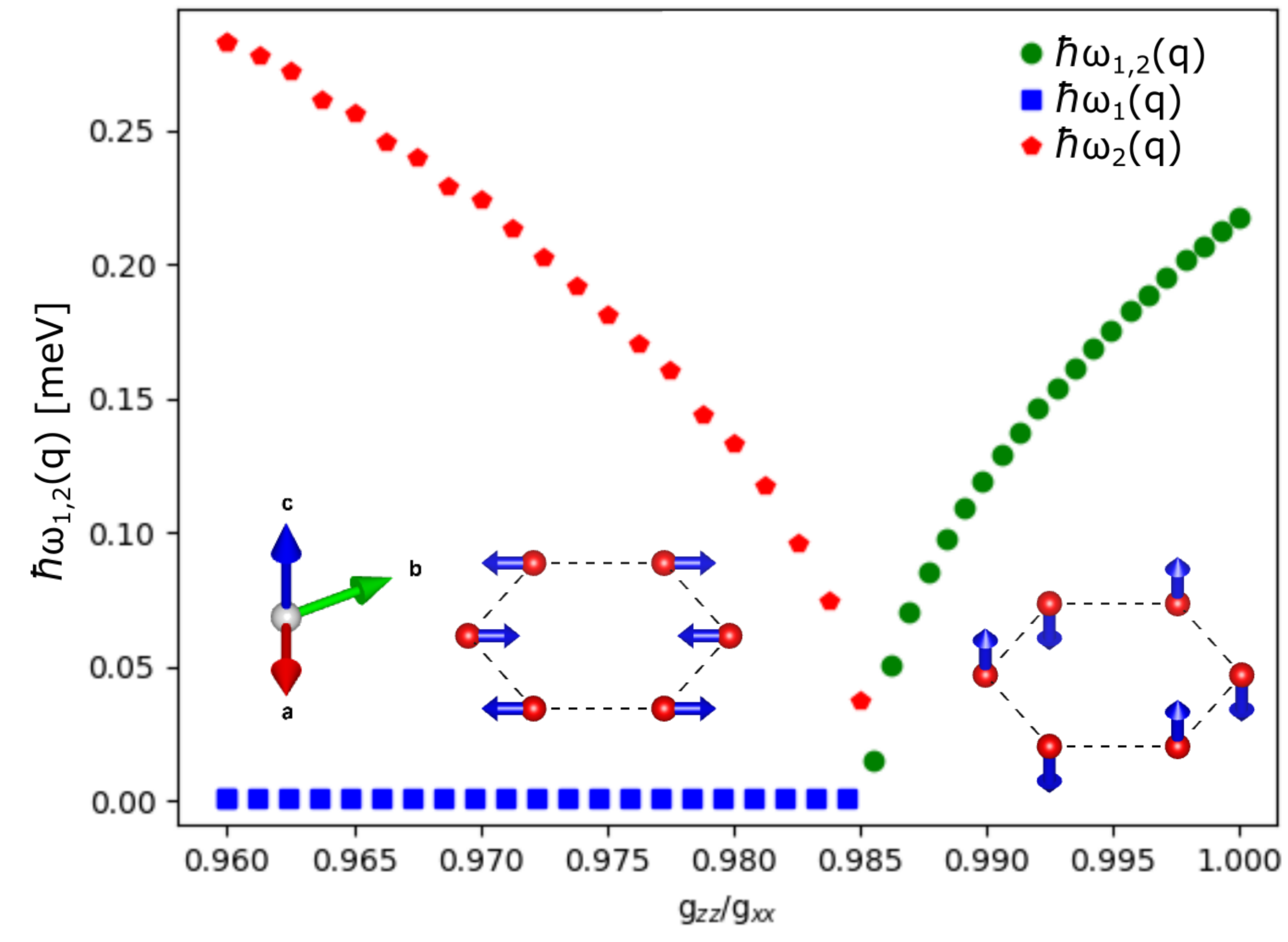}
\captionsetup{labelformat=empty}
\caption{
	\textbf{Fig. S2 $\vert$ Dependence of the energy gap as a function of easy-plane anisotropy. } 
	Above $\rm g_{zz}/g_{xx}$ = 0.985 = $\rm g_{crit}$, the calculated branches $\rm \omega_{1}(q)$ and $\rm \omega_{2}(q)$ are degenerate while for $\rm  g_{zz}/g_{xx} < g_{crit}$ the two spin-wave branches split. 
	All points are calculated with a precision of $\sim$ 0.005 meV.
	The magnetic configurations shown in the figure correspond to a N\'eel antiferromagnet with spins aligned along the c-axis for $\rm g_{zz}/g_{xx} >$ 0.985 and in the hexagonal plane for $\rm g_{zz}/g_{xx}<$ 0.985. 
 }
\label{YbBr3_Gap}
\end{center}
\end{figure}

\clearpage
\subsection*{\textsf{S3: Calculated susceptibility for a Yb$_{6}$ hexamer.}}
	
\begin{figure}[hb]
\begin{center}
\includegraphics*[width =12cm]{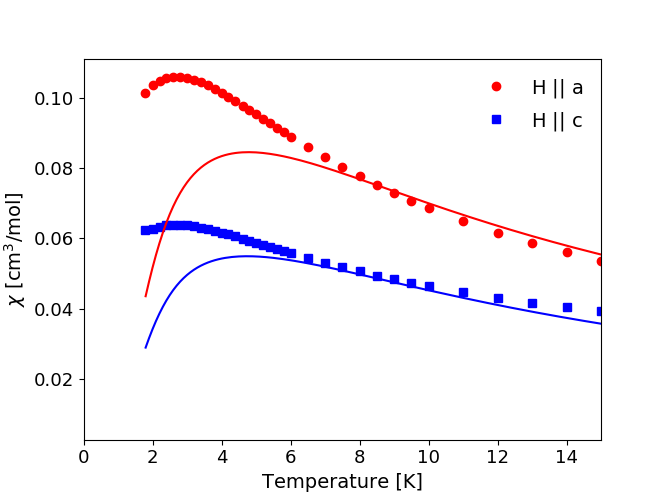}
\captionsetup{labelformat=empty}
\caption{
	\textbf{Fig. S3 $\vert$ Calculated susceptibility for a Yb$_{6}$ hexamer.} 
	The measured low-temperature magnetic susceptibility is shown together with the calculation (solid lines) for a 
	single plaquette with $S=1/2$, and the Hamiltonian $H_{\rm h} + H_{\rm dip}$ of the main text. 
 }
\label{YbBr3_HexamerCalculation}
\end{center}
\end{figure}

\clearpage

\section*{\textsf{References}}
\begingroup
\renewcommand{\section}[2]{}

\endgroup

\end{document}